\documentclass[prb,preprint]{revtex4}
\usepackage[dvips]{graphicx}
\usepackage{amsmath}
\usepackage{bm}
\usepackage{color}

\begin{document}
\title{Current-induced microwave excitation of a domain wall confined in
a magnetic wire with bi-axial anisotropy}

\author{Katsuyoshi Matsushita}
\author{Jun Sato}
\author{Hiroshi Imamura}

\affiliation{Nanotechnology Research Institute (NRI), Advanced
Industrial Science and Technology (AIST), 
 AIST Tsukuba Central 2,
Tsukuba, Ibaraki 305-8568, Japan.}

  \begin{abstract}
   We studied the current-induced magnetization dynamics of a domain
   wall confined in a magnetic wire with bi-axial anisotropy.  We showed
   that above the threshold current density, breathing-mode excitation,
   where the thickness of the domain wall oscillates, is induced by
   spin-transfer torque.  We found that the breathing-mode can be
   applied as a source of microwave oscillation because the resistance
   of the domain wall is a function of the domain wall thickness.  In a
   current sweep simulation, the frequency of the breathing-mode
   exhibits hysteresis because of the confinement.
  \end{abstract}
  \maketitle
  
  
  Recent advances in spin electronics have revealed that the current flowing
  through a magnetic nanostructure with a non-collinear magnetization
  configuration can excite magnetization dynamics.
  Current-induced microwave generation has attracted a lot of
  attention because it will be a candidate for applications in future
  wireless telecommunication technologies.
  Most studies of current-induced microwave generation have been carried
  out in magnetic multilayers
  \cite{Katine:2000,Tsoi:2000,Kiselev:2003,Rippard:2004,Covington:2004,Krivorotov:2004,Kasa:2005,Mancoff:2005}. 
  In these experiments, uniform precession
  of the free layer magnetization\cite{Slavin:2005,Rezende:2005} is driven by spin-transfer torque\cite{Slonczewski:1996,Berger:1996,Slonczewski:1999},
  and the motion of the free layer magnetization is measured using the
  CPP-GMR or TMR effect.  
  
  Only a few works have theoretically suggested on the current-induced
  microwave generation of a domain wall\cite{He:2007,Ono:2008}.  He and
  Zhang proposed an application of the oscillating motion of the domain
  wall under current and magnetic field as a source of microwave
  oscillation\cite{He:2007}.  Ono and Nakatani proposed a microwave
  oscillator using the rotating motion of the domain
  wall\cite{Ono:2008}.

   On the other hand, it is known that a domain wall produces an
   additional resistance in magnetic wires.  According to the theory of
   Levy and Zhang\cite{Levy:1997}, the resistivity of a domain wall is
   inversely proportional to the square of the domain wall thickness.
   Therefore, if we excite a breathing-mode where the thickness of the
   domain wall oscillates by application of a dc current, we can use the
   oscillation as a microwave source.  From the view point of physics,
   it is also important to study the current-induced magnetization dynamics
   (CIMD) of the geometrically confined domain wall since CIMD is in
   general different from magnetic-field-induced magnetization dynamics
   and little is known about the CIMD of a geometrically confined
   domain wall.

\begin{figure}
   \centerline{ \includegraphics[width=\columnwidth]{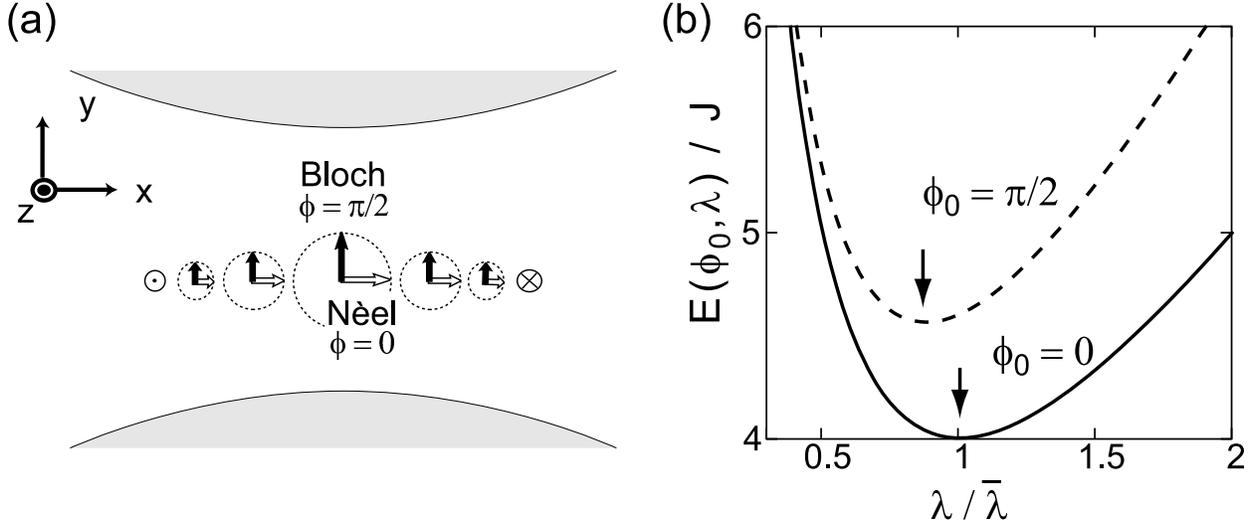} }
   \caption{ (a) Bloch and Ne{\`e}l walls confined in a magnetic wire
   are schematically shown.  The regions outside the wire are
   represented by shading.  The solid and hollow arrows represent the
   magnetization vector of the Bloch and Ne{\`e}l walls,
   respectively. The dotted circles represent the uniform rotation
   where the exchange energy of the domain wall takes a constant value.
   (b) Potential energies are plotted against the width of the domain
   wall $\lambda$ for $\phi_0=0$ (solid line) and $\phi_0=\pi/2$ (dashed
   line).  The horizontal axis is normalized by the width of the domain
   wall of the ground state for $\phi_0=0$.  The energy minima are
   indicated by arrows.}  \label{fig:model}
  \end{figure}

  In this paper, we investigated the CIMD of
  a domain wall confined in a magnetic wire with bi-axial anisotropy by a
  confining potential due to the wire shape shown in Fig.~\ref{fig:model}(a).
  We showed for a confined domain wall that above the threshold current
  density, spin-transfer torque induces breathing-mode excitation\cite{Dantas:2001,
  Thiaville:2004}, where the thickness of the
  domain wall oscillates. 
  The current-induced breathing-mode can be applied for a
  microwave oscillation because the resistance of the domain wall is a
  function of the domain wall thickness.
  We also found that if the current density is
  adiabatically changed, the frequency of the breathing-mode shows a
  hysteresis loop because the confining potential enables the
  breathing and pinning states to coexist below the threshold current
  density.
  
  The system we consider is a 180$^\circ$ domain wall confined by a potential due to the wire shape, 
  as shown in Fig.~\ref{fig:model}(a).  For simplicity, we modeled the
  system as a one-dimensional domain wall along the
  $x$-axis with a confining potential, where the directions of the
  magnetization vectors are represented as $\bm M =
  (\sin\theta\cos\phi,\sin\theta\sin\phi,\cos\theta)$, where $\theta$
  and $\phi$ denote, respectively, polar and azimuthal angles for each
  position.  We assume that the system has a bi-axial anisotropy such
  that the $z$- and $y$-axes are easy and hard axes, respectively.
  
  Let us begin with a brief introduction of the theory of a
  one-dimensional domain wall\cite{Bouzidi:1990,Braun:1996,Tatara:2004,Thiaville:2002}.  In the absence of current and confining
  potential, the energy of the system is expressed as
  \begin{equation}
   \label{eq:energy} 
   E 
   \! = \!\!
   \int_{-\infty}^{\infty}
   \!\!
   dx J\left[
   \theta'^2 \!\! + \sin^2\theta\phi'^2\right]
   \!+\! K_{\rm {\rm
   e}}\sin^2\theta
   \left[1 \!+ \! r\sin^2\phi\right],
  \end{equation}
  where the primes denote differentiation with respect to spatial $x$ coordinate.  
  The first term is the exchange stiffness energy with a stiffness
  constant of $J$.  The second term represents the bi-axial anisotropy
  characterized by the anisotropy constant for the easy axis, $K_{\rm
  e}$, and the ratio, $r = K_{\rm h}/K_{\rm e}$, where $K_{\rm h}$
  is the anisotropy constant for the hard axis.
  
  It should be noted that if the system has uni-axial anisotropy, i.~e.,
  $r=0$, the ground state is degenerate with respect to the azimuthal
  angle $\phi$ which is independent of the spatial coordinate.  Since
  the magnetization vectors of the top and bottom electrodes are aligned
  parallel to the $z$-axis, the Bloch and N\'{e}el walls correspond to
  the azimuthal angle of $\phi=\pi/2$ and $\phi=0$, respectively.  The
  magnetization vectors of the other ground states lie on the dotted
  circles as shown in Fig.~\ref{fig:model}(a).  The polar angle configuration $\theta$
  of the ground state is given by
  \begin{equation}
   \label{eq:polar} \theta(x) =
    \arccos\left[\tanh\left(\frac{x}{\lambda}\right)\right],
  \end{equation}
  where $\lambda=\sqrt{J/K_{\rm e}}$ is the domain wall thickness.  The
  degeneracy of the ground state with respect to the
  azimuthal angle $\phi$ is broken if the system has bi-axial anisotropy,
  i.e., $r \ne 0$.  Assuming that the azimuthal angle $\phi$ is
  independent of the spatial coordinate, the polar angle configuration of the minimum
  energy state for each value of $\phi$ is expressed by
  Eq.~\eqref{eq:polar} with $\lambda=\bar \lambda
  \left(1+r\sin^{2}\phi\right)^{-1/2}$, where $\bar{\lambda} =
  \sqrt{J/K_{\rm e}}$ represents the domain wall thickness of the ground
  state with $\phi=0$.

  Since we are not interested in the domain wall propagation along the
  wire but in the breathing-mode excitation,
  we assume that the domain wall is confined in a certain region
  of the wire by a confining potential $V_{\rm cf}$.
  We also assume that the characteristic length of $V_{\rm cf}$ is much larger than $\bar \lambda$.
  We adopt $\phi_0$, $X$ and $\lambda$ as collective coordinates\cite{Rajaraman:1982,Takagi:1996}.
  Here $\phi_{0}$ is defined by
  \begin{equation}
   \label{eq:phi_zero} \phi_0 =
   \frac{1}{2\lambda}\int_{-\infty}^{\infty} dx \phi\sin^2\theta,
  \end{equation}
 and $X$ is defined through the polar angle configuration of 
the ground state, $\theta_{0}$, given by
  \begin{equation}
   \label{eq:theta_zero} \theta_0 =
   \arccos\left[\tanh\left(\frac{x-X}{\lambda}\right)\right],
  \end{equation}
  where $X$ denotes the position of the domain wall center.
  Substituting Eqs. (3) and (4) into Eq. (1) as $\theta = \theta_{0} +
(\theta - \theta_{0}$) and  $\phi = \phi_{0} + (\phi - \phi_{0})$, and
ignoring the spin-wave excitation of $\theta-\theta_{0}$ and
$\phi-\phi_{0}$, the energy of the domain wall is obtained as
  \begin{equation} 
   \label{eq:ec} E(\phi_0,\lambda) = \frac{2J}{\bar \lambda} \left[
    \frac{\bar \lambda}{\lambda} + \frac{\lambda}{\bar \lambda} \left(
    1+r\sin^2 \phi_0 \right) \right]. 
  \end{equation}
  In Fig.~\ref{fig:model}(b) we plot the domain wall energy of
  Eq.~\eqref{eq:ec} with $r=0.3$ as a function of the normalized
  domain wall thickness $\lambda/\bar{\lambda}$.  The solid and dotted
  lines correspond to
  $\phi_{0}=0$ and $\pi/2$, respectively. As shown in Fig.~\ref{fig:model}(b),
 the value of $\lambda$ which minimizes the domain wall energy
  depends on $\phi_0$.  
  Therefore, if the precession of the domain wall
  around the azimuthal axis is induced by an applied current, the
  oscillation of $\lambda$ and therefore the resistance of the domain
  wall are also induced.
   
  For simplicity we
  assume that the confining potential takes the form $V_{\rm cf} =
  F_{0}{L}\left[\sinh^2(X/L)\right]$, where $F_0$ and $L$ represent
  the magnitude and the characteristic length of the confining potential.  As mentioned above we also assume
  that the characteristic length $L$ of the confining potential is much
  longer than the thickness of the domain wall $\lambda$.
  Thus the domain wall thickness is not related to $L$ 
  and  the domain wall is different from the so-called geometrically
  confined domain wall whose thickness is determined by $L$\cite{Bruno:1999}. 
   
  In order to systematically derive the equation of motion described by
  the collective coordinates, we adopt the Lagrangian
  method\cite{Tatara:2004,Thiaville:2002,Zhang:2004,Shibata:2005}.
  The Lagrangian corresponding to the torque exerted on magnetizations of the domain wall by an applied
  current\cite{Zhang:2004,Shibata:2005} is given by
  \begin{equation}
   -\int_{-\infty}^{\infty}dx\, \frac{\mu_{\rm B} P j_{\rm e}}{(1+\xi^2)\gamma e}
    \phi'\left[1-\cos\theta\right],
  \end{equation}
  were $\xi$ is the ratio between the precession time due to the exchange interaction and the spin relaxation time for spin accumulation;
 $e$ is the electric charge of an electron; $j_{\rm e}$, the charge current
  density; $P$, the spin polarization of the charge current and $\gamma$, the
  gyromagnetic constant. Then the total Lagrangian of the system under the applied
  current\cite{Thiaville:2002,Tatara:2004,Shibata:2005} is given by
  \begin{align}
   \label{eq:Lagrangian} {\cal L} &=
   \frac{1}{\gamma}\int_{-\infty}^{\infty} dx (\dot \phi - j\phi')
   \left[1-\cos\theta\right] \nonumber \\ & -J\int_{-\infty}^{\infty}
   dx \left[ \theta'^2+\sin^{2}\theta\,\phi'^2\right] \nonumber \\ &-
   K_{\rm {\rm e}}\int_{-\infty}^{\infty} dx \sin^2\theta
   \left[1+r\sin^2\phi\right] -V_{\rm cf},
  \end{align}
  where  $j = \mu_{\rm B} Pj_{\rm e}/(1+\xi^2)e$ represents the spin-current density.
  The dots denote differentiation with respect to time $t$.
  In order to obtain the effective Lagrangian described by the collective coordinates $X$, $\phi_0$ and $\lambda$, 
  we substitute $\theta_0$ and $\phi_0$ for $\theta$ and $\phi$ in
  Eq. \eqref{eq:Lagrangian}, respectively, and then perform integration
  with respect to $x$.  We obtain the following effective Lagrangian,
   \begin{eqnarray}
 {\cal L} = \frac{2}{\gamma}(\dot X + j)\phi_0 - \frac{2J}{\lambda} -2\lambda K_d\left( 1+r\sin^2\phi_0 \right) - V_{\rm cf}.\label{ eq: effective lagrangian}
 \end{eqnarray}
 
In the Lagrangian formalism, the effect of the Gilbert damping \cite{Gilbert:1955} is
conventionally taken into account by the Rayleigh dissipative function
method\cite{Landau:mechanicsSS25,Thiaville:2002}. The dissipation
function is defined by
 \begin{eqnarray}
 {\cal F} = \frac{\alpha}{2\gamma}\int dx \left\{(\dot \theta - \frac{\beta}{\alpha} j \theta')^2 + \sin^2\theta(\dot \phi - \frac{\beta}{\alpha}j\phi')^2 \right\},\label{eq: dissipation function}
 \end{eqnarray}
 where $\alpha$ is the Gilbert damping constant. The terms proportional to $\beta$ reproduces so-called $\beta$-term\cite{Zhang:2004,Barnes:2005,Thiaville:2005,Kohno:2006,Tatara:2006} and describes drift effect\cite{Seki:2008}.
 One can easily confirm that Eq.~\eqref{eq: dissipation function}
 reproduces the torques coming from the Gilbert damping and $\beta$ terms as $ - {\bm M} \times \delta {\cal F} /\delta  \dot {\bm M} = - \alpha {\bm M} \times \dot {\bm M} + \beta j {\bm M} \times {\bm M}'$.
  In order to obtain the effective dissipation function described by the
 collective coordinates $X$, $\phi_0$ and $\lambda$, in the same way as
 the derivation of the effective Lagrangian, we substitute $\theta_0$
 and $\phi_0$ for $\theta$ and $\phi$, respectively, and then perform
 integration for the coordinate $x$.
  We obtain the following effective dissipation function,
 \begin{eqnarray}
 {\cal F} = \frac{\alpha}{\gamma} \left\{ \frac{\dot X^2}{\lambda} + \lambda \dot \phi_0^2 + \frac{\pi^2}{12}\frac{\dot\lambda^2 }{\lambda}+\frac{2\beta}{\alpha\lambda}j\dot X\right\},\label{ eq: effective dissipation function}
 \end{eqnarray}
  where the terms which is independent of the time derivatives of the collective coordinates are dropped.
  
  The equation of motion for the domain wall is obtained by using
  the effective Lagrangian of Eq.~\eqref{ eq: effective lagrangian},
  the effective dissipation function of Eq.~\eqref{ eq: effective
  dissipation function} and the Euler-Lagrange equation,
\begin{eqnarray}
 \frac{\partial}{\partial t}\frac{\delta  {\cal L}}{\delta \dot q} = \frac{\delta  {\cal L}}{\delta q} - \frac{\delta  {\cal F}}{\delta \dot q},
 \end{eqnarray}
 where $q$ being in $X$, $\phi_0$ and $\lambda$. After some algebra we obtain 
  \begin{align}
   &\dot \phi_0 + \alpha \frac{\dot X}{\lambda} = - \frac{\beta}{\lambda}j
   + \gamma F_{\rm cf}^X(X), \label{eq: eq of motion on phi_0}\\ & -
   \dot X + \alpha{\bar \lambda}\dot\phi_0 \frac{\lambda}{\bar \lambda}
   = j - \gamma \frac{rJ}{\bar\lambda^2} \lambda\sin2\phi_0, \label{eq:
   eq of motion on X}\\ &\frac{\pi^2\alpha}{12\gamma}\dot \lambda =
   \frac{J}{{\bar\lambda}}\left[\frac{\bar \lambda}{\lambda} -
   \frac{\lambda}{\bar\lambda}\left(1+ \frac{r}{2}(1-\cos 2\phi_0)\right)\right],
   \label{eq: eq of motion on l}
  \end{align} 
  where $F_{\rm cf}^X= - \partial_X V_{\rm cf}$. 
  Equations \eqref{eq: eq of motion on phi_0},  \eqref{eq: eq of motion
  on X} and   \eqref{eq: eq of motion on l} reduces to Slonczewski's
  equation of domain wall motion\cite{Slonczewski:1972,Hubert:1998} when
  the dissipation for dynamics of $\lambda$ is
  neglected.
  The $\beta$-proportional term in
  Eq.~\eqref{eq: eq of motion on phi_0} represents the torque from the
  so-called $\beta$-term of the Landau-Lifshitz
  equation\cite{Landau:1935}.
  We note that $\lambda$ is a dynamical variable and is independent of
  $\phi_0$ in Eqs.~\eqref{eq: eq of motion on phi_0}-\eqref{eq: eq
  of motion on l}\cite{comment_for_dynamics_of_lambda}.

  We performed numerical simulations based on Eqs.~\eqref{eq: eq of
  motion on phi_0}-\eqref{eq: eq of motion on l}. The equations were
  solved using the implicit Runge-Kutta method.  Since the
  confinement force, $F_{\rm cf}^X$, is much
  larger than the $\beta$-term in the present situation, we 
  ignored the first term of the right-hand-side of Eq.~\eqref{eq: eq of
  motion on phi_0}.
  We took the parameters of the
  confining potential to be $L=2.0\bar\lambda$, and $F_0=100.0J/\bar\lambda^2$ to efficiently confine the domain wall to the
  confinement region. This condition corresponds to the case in which the
  cross section of the wire exponentially increases tenfold in the linear
  dimension by the displacement to $X=L$ from $0$. We also set the
  Gilbert damping constant at 0.01 to reproduce typical experimental
  systems.  At initial time $t=0$, the thickness, position, and
  azimuthal angle were taken to be $\lambda=\bar \lambda$,$X=0$, and
  $\phi_0=0$, respectively.

  \begin{figure}[tb]
   \centerline{ \includegraphics[width=\columnwidth]{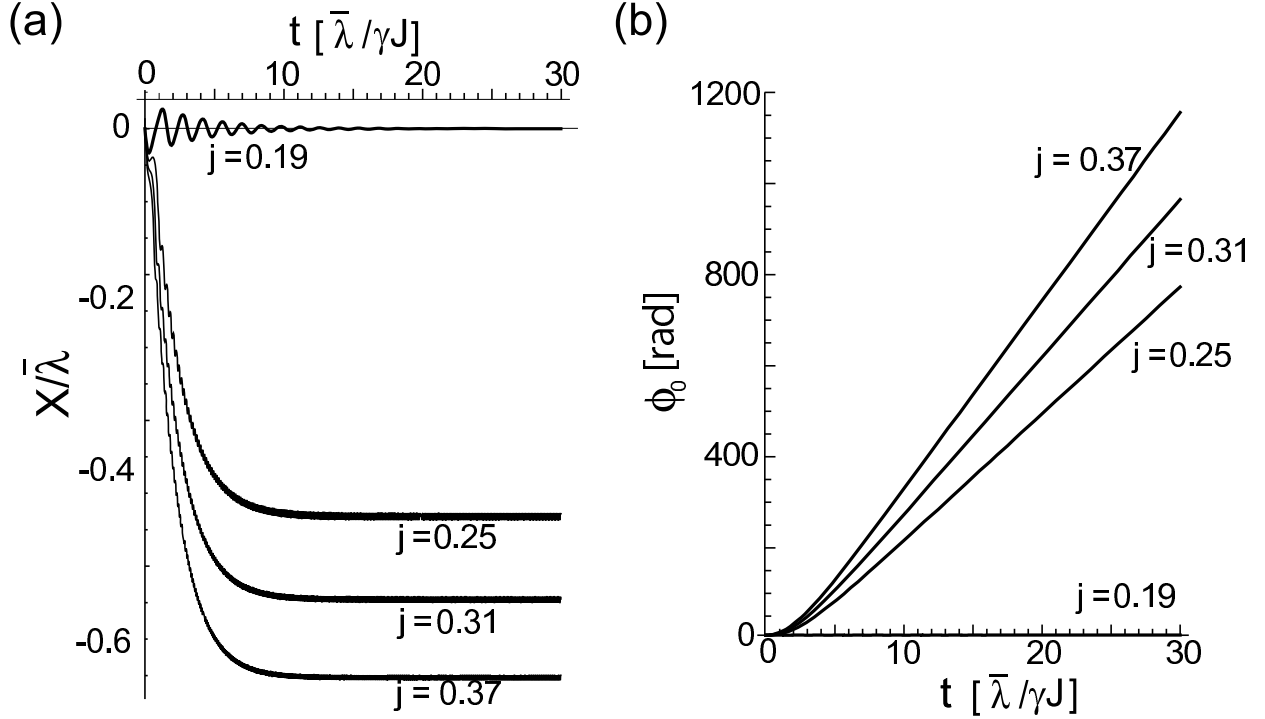} }
   \caption{ (a) The position of the domain wall $X$ is plotted as a
   function of time $t$.  The current densities are taken to be
   $j=$0.37, 0.31, 0.25 and 0.19 for top to bottom. The unit of the
   current density is taken to be $\gamma J/\bar{\lambda}$.  (b) The
   angle $\phi_0$ is plotted as a function of time $t$.  The plot for
   $j=$0.19 lies on the horizontal axis.  In both panels, the unit of
   time is taken to be $\bar{\lambda}^{2}/(\gamma J)$.  }
   \label{fig:fig2}
  \end{figure}

  In Figs.~\ref{fig:fig2}(a) and \ref{fig:fig2}(b) we plot the position
  $X$ and the angle $\phi_0$ of the domain wall as a function of time
  for various values of $j$, respectively.  The current is switched on
  at $t=0$ and then is kept fixed.  As long as the spin-current
  density $j$ is smaller than the critical value $j_{\rm c}$, the position
  of domain wall $X$ showed little deviation from its initial value of $X =
  0$.  In our simulation $j_{\rm c} = 0.22$ $\gamma J/\bar \lambda$. 
  Hereafter, the unit of the spin-current density is taken to be $\gamma
  J/\bar \lambda$.
  As shown in Fig.~\ref{fig:fig2}(b),
  the angle $\phi_0$ also shows little deviation from its initial value
  $\phi_{0}=0$ for $j=0.19 <j_{\rm c}$.  Above the critical current, $j
  \ge j_{\rm c}$, $X$ moves to a certain position which is determined by
  the competition between the confining potential and the spin-transfer
  torque.  The angle $\phi_0$ linearly increases with increasing time
  and the domain wall precesses around the $x$-axis.
  
  The depinning of the angle $\phi_0$ shown in Fig.~\ref{fig:fig2}(b)
  is similar to Walker's breakdown \cite{Hubert:1998,Tatara:2004}.
  However, the $j_{\rm c}$ we obtained is not equal to Walker's threshold
  $j_{\rm W}$ =  $rJ/\bar \lambda$. As we shall show later, the
  difference in $j_{\rm c}$ and $j_{\rm W}$ reflects the essential
  difference in dynamics between confined and unconfined domain walls,
  such that coexistence of the oscillation and pinning states for $j <
  j_{\rm W}$ depends on the confining potential.

\begin{figure}[tb]
   \centerline{ \includegraphics[width=\columnwidth]{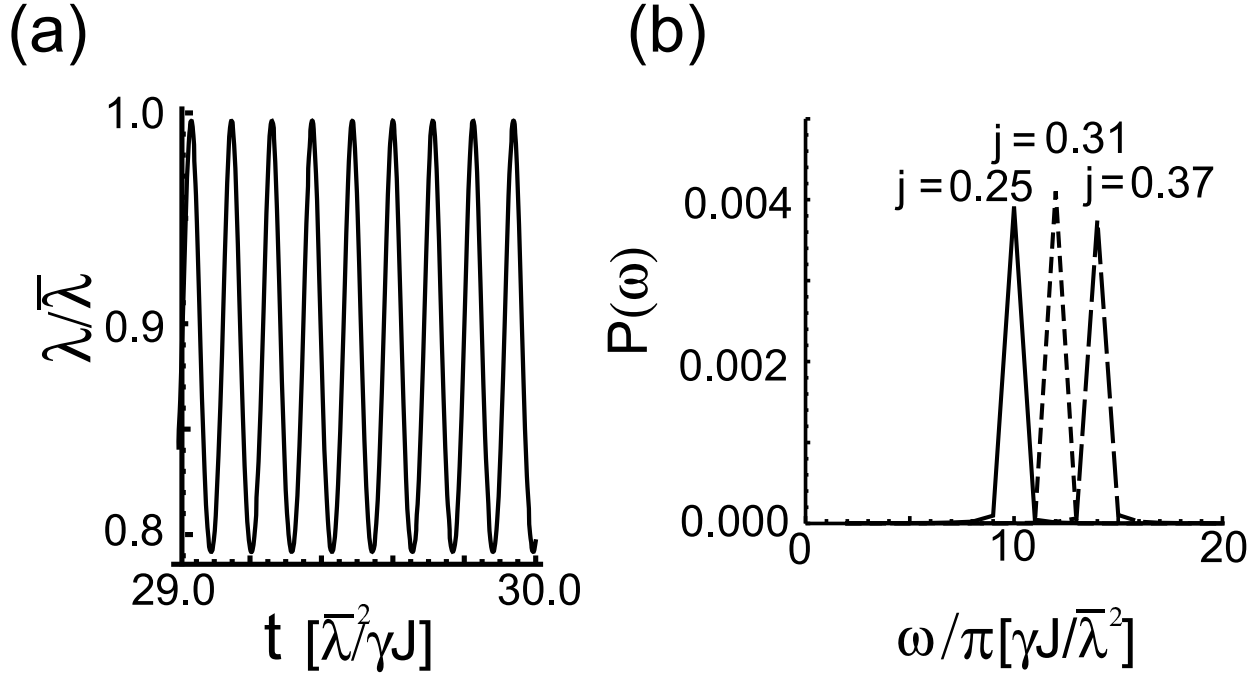} }
   \caption{Thickness oscillation. (a) The normalized width of the domain wall
   $\lambda/\bar{\lambda}$ for j=0.25 is plotted as a function of time $t$.  (b)
   The power spectrum density $P(\omega)$ of magnetoresistance obtained
   by using the theory of Levy and Zhang\cite{Levy:1997}.  Solid, dotted
   and dot-dashed lines correspond to the current density of 0.25, 0.31,
   and 0.37, respectively.   The unit of the current density is taken to
   be $\gamma J/\bar{\lambda}$.
   } \label{fig:fig3}
  \end{figure}

  Since the system has bi-axial anisotropy, the precession of the
  domain wall induces the oscillation of the domain wall thickness $\lambda$, the breathing mode\cite{Dantas:2001}, as shown in Fig.~\ref{fig:fig3}(a).  According to Levy and Zhang's
  theory\cite{Levy:1997}, the resistance $\Delta R$ of a domain wall
  depends on its thickness as $\Delta R \sim 1/\lambda$.  Thus, the
  resistance $\Delta R$ oscillates due to the breathing mode.

  In Fig.~\ref{fig:fig3}(b) we plot the power spectrum density of
  $(\bar{\lambda}/\lambda) \sim \Delta R$ defined as
  \begin{equation}
   P(\omega) = \int d\tau e^{-i\omega\tau}\int dt
    \frac{\bar\lambda^2}{\lambda(t)\lambda(t-\tau)}.
  \end{equation} 
  For $j > j_{\rm c}$, the oscillation in $\Delta R$ was observed as
  expected above.   For the typical experimental situation the frequency is on the order of
  several tens GHz \cite{Ono:2008}.
  For each value of $j$, $P(\omega)$ has a
  single sharp peak.  This means that the system is a useful candidate for
  a microwave source.  The peak frequency of the power spectrum is
  proportional to $j$, which means that we can control the microwave
  frequency by the current. 
   
  The intensity at the peaks is on the order of 0.01 and depends on 
  the amplitude of the thickness oscillation.  In this case the
  amplitude of the thickness oscillation is on the order of
  $0.1\bar{\lambda}$.  The value of the intensity is consistent with the
  amplitude of the thickness oscillation because $P(\omega)$ is
  proportional to the square of the amplitude normalized by
  $\bar\lambda$.  The amplitude of the oscillation is about
  $\bar{\lambda}r$.  Thus, the intensity is
  controlled by the ratio of anisotropy constants, $r$,  and does not
  depend on $j$.
We note that $\lambda$ is always shorter than $\bar \lambda$ as shown in Fig.~\ref{fig:fig3}(a). Namely $\lambda$ does not agree with the thickness minimizing Eq.\eqref{eq:ec} because of the effect of the Gilbert damping in Eq.~\eqref{eq: eq of motion on l}\cite{comment_for_dynamics_of_lambda}.

\begin{figure}[tb]
   \centerline{ \includegraphics[width=0.75\columnwidth]{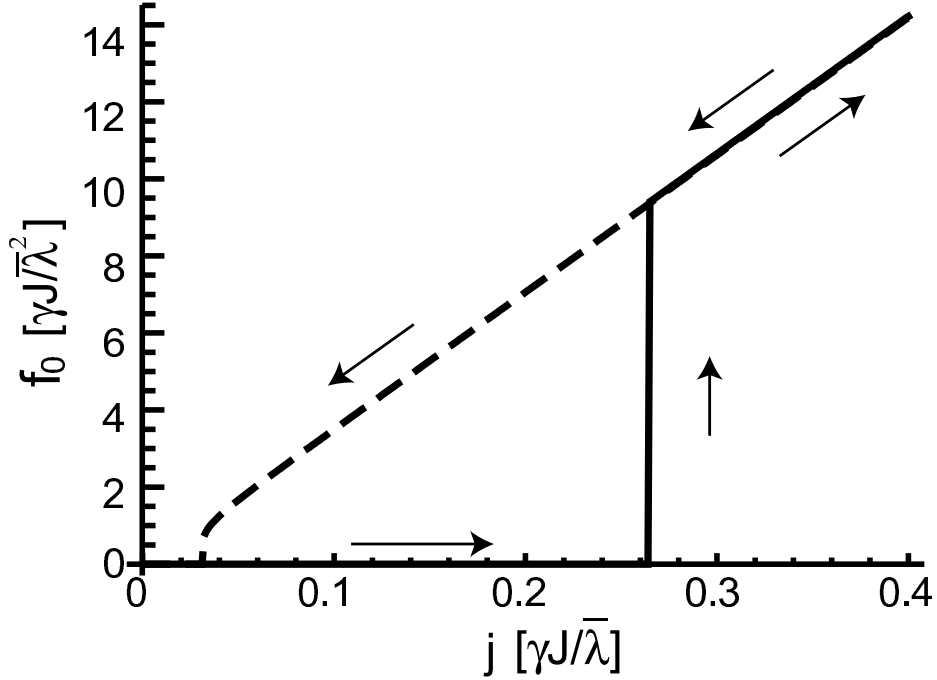} }
   \caption{ The breathing-mode frequency $f_{0}$ is plotted against the current density $j$.  A hysteresis
   loop appears with increasing and decreasing applied current
   density (indicated by arrows).  } \label{fig:fig4}
  \end{figure}

  Next we move onto adiabatic current sweeping. 
  The current was adiabatically increased from 0 to 0.4
  which is above Walker's threshold current $j_{\rm W}$, and then was
  adiabatically decreased from 0.4 to 0.
  From Eq.~\eqref{eq: eq of motion on l}, the frequency of the
  breathing mode in the steady state is related to the velocity of $\phi_0$ as $f_0 \equiv \dot \phi_0/\pi$.
  Figure~\ref{fig:fig4} shows the breathing-mode frequency $f_0$.
  As the current increased from 0, the frequency was kept at zero below the
  threshold current $j'_{\rm c} = j_{\rm W}\lambda_{\rm min}/\bar
  \lambda$, where $\lambda_{\rm min} \equiv \min \lambda =
  \bar\lambda\sqrt{1-r}$.  When
  the current reached $j=j'_{\rm c}$ in the simulation, the frequency jumped 
  to a certain value and then linearly increased as
  the current further increased up to 0.4.
  As the current decreased from 0.4, the frequency linearly
  decreases down to a much lower current than $j'_{\rm c}$. 
  That is to say, below the threshold current, the breathing and pinning
  states coexist.
  The behavior differs surprisingly from that $\left<\dot \phi_0\right> \sim
  \sqrt{j^2-j_{\rm W}^2}$ \cite{Tatara:2004} above the threshold
  current $j_{\rm W}$ in Walker's theory, where a unique state is
  permitted for each current.

  The reason for current-dependence behavior of the breathing
  motion frequency is that the confining potential enables the breathing
  and pinning states to coexist.  In fact we can see two time-averaged
  solutions of Eqs.~\eqref{eq: eq of motion on phi_0} - \eqref{eq: eq of
  motion on l} : the pinning solution is $<X>=<\dot X>=<\dot \phi_0>=0$,
  $<\phi_0> \sim \arcsin(j\lambda/\gamma rJ)/2$ and the breathing
  solution is $<\dot X>=0$, $<\dot \phi_0> \sim
  <j/\alpha\lambda> \sim
  \gamma<F_{\rm cf}^{X}(X)>$. 
  As mentioned before, the 
  threshold $j_{\rm c}$ originates from the coexistence of the two states.
  We note that if the confining
  potential does not exist the breathing mode vanishes because $<F_{\rm
  cf}^{X}(X)>=0$. The existence of the confining potential induces a
  drastic effect in the domain wall motion.

  In conclusion, we examined the current-induced magnetization dynamics of a
  domain wall confined in a magnetic wire with bi-axial anisotropy. 
   We showed that breathing-mode excitation, which produces resistance
  oscillation,  is induced by spin-transfer torque.
   The result means that the confined domain wall is a powerful candidate
   for a microwave oscillator. 
   We also found that the dependence of the frequency of the
   breathing mode on the current shows a characteristic hysteresis
   loop originating from the confining potential.

  The authors thank M.~Doi, H.~Iwasaki, M.~Ichimura, K.~Miyake, M.~Takagishi,
  M.~Sahashi, M.~Sasaki, T.~Taniguchi, N.~Yokoshi and K.~Seki for useful
  discussions.  The work was supported by NEDO and MEXT.Kakenhi(19740243).



\begin{thebibliography}{37}
\bibitem[{com()}]{comment_for_dynamics_of_lambda}
\bibinfo{note}{One may consider that the oscillation of the domain wall
  thickness we obtained can be derived by minimizing Eq.~\protect\eqref{eq:ec}
  in quasistatic limit, $\lambda=\bar \lambda
  \left(1+r\sin^{2}\phi\right)^{-1/2}$, even without solving Eq.~\eqref{eq: eq
  of motion on l}. However, as shown in Fig.\ref{fig:fig3}(a), $\lambda$ is not
  equal to $\bar \lambda \left(1+r\sin^{2}\phi\right)^{-1/2}$ in general. The
  difference between the oscillation of the domain wall thickness obtained by
  solving Eq.~\eqref{eq: eq of motion on l} and that of $\lambda=\bar \lambda
  \left(1+r\sin^{2}\phi\right)^{-1/2}$ is essential for the system under high
  current density. For example, let us consider the case where the time scale
  of the damping of $\lambda$ is larger than that of the collective precession
  of $\phi_0$. Such a situation is realized by the large spin-transfer torque
  under high current density. In such a case $\lambda$ cannot relax to $\bar
  \lambda \left(1+r\sin^{2}\phi\right)^{-1/2}$ and we have to solve
  Eq.~\eqref{eq: eq of motion on l} without qusistatic approximation.}
\end{thebibliography}
\end{document}